# Colloidal Deposit of an Evaporating Sessile Droplet on a Non-uniformly Heated Substrate

Laxman K. Malla[a], Rajneesh Bhardwaj[b, *], Adrian Neild[c]

[a] IITB-Monash Research Academy, Indian Institute of Technology Bombay, Mumbai, 400076, India.

[b] Department of Mechanical Engineering, Indian Institute of Technology Bombay, Mumbai, 400076 India.

[c] Department of Mechanical and Aerospace Engineering, Monash University, Melbourne, VIC 3800, Australia.

*Corresponding author (email: rajneesh.bhardwaj@iitb.ac.in)

*Abstract*

The pattern and profile of a dried colloidal deposit formed after evaporation of a sessile water droplet containing polystyrene particles on a non-uniformly heated glass are investigated experimentally. In particular, the effects of temperature gradient across the substrate and particles size are investigated. The temperature gradient was imposed using Peltier coolers, and side visualization, infrared thermography, optical microscopy, and optical profilometry were employed to collect the data. On a uniformly heated substrate, a ring with an inner deposit is obtained, which is attributed to axisymmetric Marangoni recirculation and consistent with previous reports. However, the dimensions of the ring formed on a non-uniformly heated substrate are significantly different on the hot and cold side of the substrate and are found to be a function of the temperature gradient and particles size. In the case of smaller particle size, the contact line on hot side depins and together with twin asymmetric Marangoni recirculations, it results in a larger ring width on the cold side as compared to the hot side. In contrast, the contact line remains pinned in case of larger particles, and the twin asymmetric Marangoni recirculations advect more particles on the hot side, resulting in a larger ring width at the hot side. A mechanistic model is employed to explain why the depinning is dependent on the particle size. A larger temperature gradient significantly increases or decreases the ring width depending on the particle size, due to a stronger intensity recirculation. A regime map is proposed for the deposit patterns on temperature gradient-particle size plane to classify the deposits.

# 1 Introduction

Self-assembly of colloidal particles in an evaporating sessile droplet on a solid surface is a much-studied problem in interface science in the last two decades [1] and is also known as the coffee-ring effect. A schematic of a sessile droplet containing colloidal particles is shown in Figure 1(a). As pointed out by Goldstein [2] recently, Brown [3] was the first researcher to visualize the advecting particles towards the contact line in an evaporating droplet on a solid surface. There is a large body of work reported for self-assembly of colloidal particles in evaporating thin films (e.g. Ref. [4]), that provided the context to this problem. In a seminal paper, Deegan et al. [5] explained that the capillary flow inside the droplet (bulk flow) caused by non-uniform evaporation of a sessile droplet is the reason of formation of a ring-like colloidal deposit after the evaporation. The evaporation mass flux is the largest near the contact line and the bulk flow is radially outward i.e. towards the contact line [5], as shown in a schematic in Figure 1(b).

The coffee-ring effect is pivotal to designing several technical applications. As discussed in previous reviews [6, 7], applications include inkjet printing, manufacturing of bioassays, medical diagnostic techniques, patterning surfaces by colloidal particles in photonics, fabrication of electronics circuit, conductive coatings, porous films, etc. The final deposit pattern and shape may be crucial to design an application. For instance, a uniform deposit is desired in inkjet printing, as compared to a ring-like deposit. In order to engineer the deposit pattern and shape, previous studies have reported the effect of several parameters on final deposit pattern such as particles size, particles shape, concentration, hydrophobicity of the particles, pH of the colloidal suspension, substrate temperature, substrate wettability, relative humidity, electrowetting, etc (see discussion notable reviews [6–11]). One important parameter in this arena is substrate temperature that has gained significant attention since 2015 [12–14]. The focus of the present work is a non-uniformly heated substrate, that could help to design innovative applications. For example, droplet migration caused by Marangoni stresses on a surface with a temperature gradient is useful for developing microfluidic or lab-on-a-chip technologies, such as one demonstrated in Ref. [15]. In the following, previous studies that tackled the coffee-ring effect on a uniformly non-heated and uniformly heated substrate have been discussed. In order to provide context to the non-uniformly heated substrate, previous studies on evaporation of droplets of pure liquid on such a substrate have also been discussed.

On a *non-heated substrate*, the colloidal suspension which exhibits Marangoni or thermocapillary stresses (e.g., alcohol-based suspension) results in an inner deposit [16–19]. The temperature of the liquid-gas interface near the contact line is higher than the temperature on the interface at the apex of the droplet. This temperature gradient across the interface causes a surface tension gradient, which induces the Marangoni stresses and thermocapillary (interfacial) flow at the liquid-gas interface. Consequently, this flow together with the radially outward bulk flow results in an axisymmetric Marangoni flow recirculation inside the droplet (Figure 1(c)). The colloidal particles are advected towards the droplet center due to the recirculation, leading to a central bump like deposit [16, 18]. Nguyen et al. [20] reported the formation of ring-like inner deposits due to initial depinning and later pinning of the contact line. The radius of such deposit is smaller than the initial wetted radius [20]. Recently, Weon and Je [21] reported fingering-like ring patterns for a bi-dispersed suspension of 0.1 $\mu$m and 1 $\mu$m particles due to the competition between the outward radial flow and the inward Marangoni flow. Similarly, Trybala et al. [22] reported a spot-like and ring-like dried patterns for the evaporation of aqueous solutions of carbon and $TiO_2$ nanoparticles on a polyethylene film, respectively. The former pattern was attributed to the Marangoni recirculation.

In the context of a *uniform heated substrate,* Girard et al. [23] mapped the time-varying liquid-air interface temperature of an evaporating *pure water droplet* using infrared thermography. They reported that the temperature difference between the contact line and the droplet apex decreases with time. Regarding colloidal suspensions, Li et al. [12] reported the deposits obtained after evaporation of an aqueous droplet containing 0.25% (v/v) polystyrene nanoparticles on a glass substrate, heated from 30ºC to 80ºC. They reported that the "coffee-ring" changes to a "coffee-eye" due to inward axisymmetric Marangoni flow. Parsa et al. [13] reported the formation of a dual-ring pattern at the elevated substrate temperatures of 47, 64, and 81 °C due to the ring-like cluster built-up of the particles on the liquid-air interface, owing to the Marangoni flow. They reported that a further increase in the temperature to 99ºC led to the formation of multiple rings due to the stick-slip motion of the contact line. Similarly, Zhong and Duan [24] also reported the formation of dual-ring patterns with the same mechanism for the evaporation of aqueous droplets containing graphite nano-powders on silicon wafers, heated at 50ºC. In a follow-up study [25], they reported the suppression of dual ring at 84ºC due to the enhanced outward radial flow in comparison to the Marangoni flow. Patil et al. [14] studied the deposition patterns of polystyrene particles on glass and silicon wafers heated from 27 °C to 90ºC at different particles concentrations in an aqueous droplet. They reported the thinning of the ring width with an increase in the substrate temperature. In a follow-up study, Patil et al. [26] reported that with an increase in the substrate temperature, the receding angle of an aqueous droplet containing smaller polystyrene particles ($d = 0.1~\mu$m) reduces on a silicon wafer, which helps in pinning of the contact line and formation of a thin ring along with an inner deposit due to the Marangoni recirculation. Regarding bi-dispersed colloidal suspensions, Parsa et al. [27] reported the self-sorting of 1.0 $\mu$m particles towards the outer edge of the ring in the evaporation of a bi-dispersed solution containing 1.0 and 3.2 $\mu$m particles on a uniform heated substrate. Similarly, the self-sorting of the smaller particles near the contact line was reported by Patil et al. [26] during the evaporation of a bi-dispersed solution containing smaller (0.1 and 0.46 $\mu$m) and larger particles (3.0 $\mu$m) on a heated silicon wafer.

Regarding *non-uniformly heated substrate*, several studies reported migration of a *pure liquid* droplet on the surface with low contact angle hysteresis (CAH) and attributed this migration to Marangoni stresses. Brzoska et al. [28] reported migration of a PDMS droplet from hot to cold side on a silanized Si wafer with a CAH of around 2º. The temperature gradient on the substrate was varied from 0.35 to 1.08ºC/mm. Notably, they reported that the internal flow direction is from the hot side to the cold side at the liquid-gas interface. Recently, Ouenzerfi and Harmand [29] observed an opposite migration of water-3% butanol binary droplet, i.e., from the cold side to the hot side, on a substrate with 0.5ºC/mm gradient. Authors attributed it to an increase in surface tension of the binary liquid with temperature, beyond a critical value of the temperature. In contrast, the droplet does not migrate on a non-uniformly heated surface with larger CAH. Recent studies have measured the flow-field in an evaporating droplet on such a surface. Pradhan and Panigrahi [30] measured the flow field inside pure water microliter droplets evaporating on a hydrophobic siliconized coverslip. The imposed temperature gradient on the substrate was 1.8ºC/mm. They reported twin Marangoni recirculation, in which the flow direction is from the hot side to the cold side at the liquid-gas interface, in a droplet with a wetted diameter of 1.9 mm. Similarly, such twin recirculations were visualized through infrared thermography by Askounis et al. [31] in an evaporating water droplet on a copper substrate, locally heated by a laser at a point near the contact line.

A brief literature review shows that while the effect of the temperature gradient across the substrate on thermocapillary migration and internal flow fields inside *pure liquid* droplets is well-established, there are no reports of the evaporation of the droplets of *colloidal suspensions* on a non-uniformly heated substrate, to the best of our knowledge. Therefore, the primary objective of the present work is to investigate the pattern and profile of dried colloidal deposits on the non-uniformly heated substrate. In particular, the effect of particle size and intensity of temperature gradient on the deposits are presented. The present work utilizes a system with large contact angle hysteresis in order to avoid droplet migration.

## 2  Experimental details
### 2.1  Generation of droplets of colloidal suspensions on glass

Aqueous colloidal suspensions were procured from Sigma Aldrich Inc. These 10% v/v suspensions contain uniformly dispersed polystyrene latex particles (beads). Three solutions with different particles diameter, $d$ = 0.1 (LB1), 1.1 (LB11), and 3.0 μm (LB30), were used in the present study. The suspensions were used as procured from the manufacturer and were diluted by deionized water (resistivity of 18.2 MΩ·cm) to prepare a solution of $c$ = 0.1 %v/v. Once dilution was complete, solutions were sonicated for about 30 minutes to ensure a highly disperse sample. A particles concentration of 0.1% v/v is used for all measurements presented in section 4.

The properties of these suspensions are available through the manufacturer product information sheet [32] and are briefly given as follows. The standard deviation of the diameter of the particles in the solution is on the order of 5-15% of the mean diameter. The density of the particles is around 1005 kg/m$^3$, implying they are neutrally buoyant and do not sediment on the substrate. The particles are hydrophilic and the contact angle, when attached to the air-water interface, is around 43º [33]. The surface charge density of the 1.1 μm particles in an aqueous solution is around -2 μC/cm$^2$, as reported in previous measurements [34].

Droplets of volume $1.0 \pm 0.3$ μL were generated using a micropipette (Prime, Biosystem Diagnostics Inc.) and were gently placed on the substrate. Pre-cleaned glass slides (Sigma Aldrich Inc., S8902 [35]) with dimensions of 75 x 25 x 1 mm$^3$ served as the substrate. The glass slide was washed with isopropanol and was allowed to completely dry in the ambient conditions for a few minutes before the droplet was deposited on it. A fresh glass slide was used to repeat or to perform a new experiment. The roughness of the slide was measured using AFM (MFP-3D Origin, Asylum/Oxford Instruments Inc.) in non-contact mode and root mean square value ($R_a$) is around 2.9 nm. The surface potential of the glass surface is around -40 to -15 mV, as reported in previous measurements [36, 37].

### 2.2  Non-uniform heating of the substrate

In order to impose a temperature gradient on the glass slide, two Peltier coolers (TEC-12706) were used, which were maintained at different temperatures by a microprocessor-based controller (TEC-1122-SV, Meerstetter Engineering GmbH, Switzerland), as schematically shown in Figure 2. The temperature controller utilized a DC power supply (Tektronix Inc, USA, 0-32 V, 6A). A thin copper plate of 0.4 mm thickness was mounted between the glass slide and the Peltier cooler using thermal paste. An aluminum block with fins served as a heat sink beneath the Peltier coolers. A similar approach was employed by Pradhan and Panigrahi [30].

## 2.3 Thermal imaging

The temperature field on the top of the substrate was measured using an infrared (IR) camera (A6703sc, FLIR Systems Inc.), with a 25 mm (f/2.5) IR lens (Figure 2). For the droplet, the measured temperature is at the liquid-gas interface temperature since water is generally opaque to the infrared radiation. The working distance, pixel resolution, and fps of the IR camera were 22 cm, 15.3 $\mu$m per pixel and 50, respectively. The emissivity of water and glass were taken as 0.97 and 0.92, respectively, for the estimation of the temperature field by IR camera.

Three cases of the temperature gradient ($dT/dX$) prescribed on the surface were considered. The lower temperature ($T_L$) was maintained at 25ºC, and the larger temperature ($T_H$) was maintained at either 60, 50, or 40ºC to obtain $dT/dX$ = 4.2, 2.8, or 1.7ºC/mm, respectively. Measured isotherms on the surface are shown for these three cases in Figure 3(a-c). The temperature profiles along $X$ at $Y$ = 3 mm are plotted in Figure 3 (d), along with their linear fits ($R^2 \sim 0.97$). The droplet was placed approximately at ($X$, $Y$) = (3 mm, 3 mm) and the approximate temperature at this location before the droplet deposition was around 36, 45, and 51ºC for $dT/dX$ = 1.7, 2.8, or 4.2ºC/mm, respectively. The substrate was also uniformly heated at $T_s$ = 60ºC, to act as a control, using a digital controlled hot plate (15956-32, Cole-Parmer Inc.).

## 2.4 Side-visualization and optical microscopy

During the experiments, the side view of the sessile droplet was visualized using a camera (IDT Inc., MotionPro, Y-3 classic) and long-distance working objective (Qioptiq Inc.), with a white LED lamp acting as a backlight source, as shown schematically in Figure 2. The working distance and magnification of the high-speed camera were 9.5 cm, and 14 $\mu$m per pixel, respectively, and the videos were recorded at 50 frames per second (fps). Dried patterns were visualized from the top by an optical microscope (Olympus BX53F, with a magnification of 4X to 40X) for the zoomed-in view.

The initial static and the receding contact angles for different cases of d$T$/d$X$ and particle diameter ($d$) were measured from the images of the side view, assuming droplet shape as a spherical cap and are provided in Table 1. The spherical cap assumption is justified since the wetted radius is smaller than the capillary length of water and the imposed temperature gradient does not deform the liquid-gas interface, as confirmed from the side visualization of the evaporating droplet. While the measured angles do not show significant variation with d$T$/d$X$, there is a slight increase in the angles as the particle size ($d$) increases. This data is consistent with our previous study [38] in which a weak dependence of the static contact angle on the particle size was reported for a droplet of aqueous colloidal suspension on a non-heated glass. All experiments in the present study were performed three times to ensure repeatability. The ambient temperature and relative humidity were 25 ± 2 °C and 35 ± 5%, respectively.

## 3   A mechanistic model for the depinning of the contact line

A first-order model, developed in previous studies [37, 39–41], is utilized to predict depinning or pinning of the contact line against the present measurements. The model accounts for the several forces acting on the particles stacked near the contact line, shown in a schematic in Figure 4. The surface tension force ($F_s$) acting on the particle near to contact line and touching the liquid-gas interface pulls the particle away from the contact line while drag force ($F_d$) push them towards the contact line. The normal forces ($F_a$) acting on the particles are van der Waals and electrostatic forces. Here, gravity is ignored for neutrally buoyant polystyrene particles. The expressions of these forces are provided in the supporting information. The dimensionless net force ($\Delta F^*$) acting on the outermost particle along the inward horizontal direction is given by [26, 37, 39–41],

$$\Delta F^* = \frac{1}{\gamma_{LG} d} \left[ F_s \sin\theta_{rec} - \left\{ f \left( F_s \cos\theta_{rec} + nF_a \right) + nF_d \right\} \right] \qquad (1)$$

where $\gamma_{LG}$, $d$, $\theta_{rec}$, $f$ and $n$ are surface tension, particle diameter, receding contact angle, the friction coefficient between the particle and substrate in liquid medium and number of particles stacked at the contact line, respectively. The sign on the net force determines if the contact line pins ($\Delta F^* < 0$) or depins ($\Delta F^* > 0$). The values of the parameters required for the calculation of the forces are taken from previous studies [26, 37] and are provided in the supporting information. The measured values of the receding angles ($\theta_{rec}$) for different cases of the temperature gradient (d$T$/d$X$) and particles size ($d$) are listed in Table 1.

# 4  Results and discussion

Results are presented for the evaporation of 1.0 ± 0.3 $\mu$L water droplet containing polystyrene particles on a hydrophilic glass substrate for three cases of particle diameter ($d$ = 0.1, 1.1, and 3.0 $\mu$m). The hydrophilic glass substrate is maintained at a uniform temperature ($T_s$ =25 and 60°C) and a non-uniform temperature with three cases of the temperature gradient, ($dT/dX$ = 1.7, 2.8, and 4.2°C/mm).

## 4.1  Uniform temperature substrate

Optical microscopy images of deposition patterns on the substrate at $T_s$ =25, and 60°C, with three cases of particle diameter, $d$ = 0.1, 1.1, and 3.0 $\mu$m are shown in Figure 5. The patterns for each case are presented for two runs. All patterns exhibit a ring-like deposit and coupled transport phenomena during the formation of the ring is well-documented in the literature [5, 42]. The corresponding ring profiles measured using optical profilometer for the first run at $T_s$ = 25°C is shown in Figure 6 (a). The ring profiles plotted on both the left and right sides of the ring are very similar, and the profiles indicate a partial torus-like shape. The measured cross-section of the ring is consistent with our previous study [38]. Zoomed-in views of the two sides plotted in Figure 5 (a) are qualitatively consistent with the measured profiles. The ring dimensions (width and height) are almost on the same order on the two sides. Cracks in the ring were recorded at $d$ = 0.1 $\mu$m (Figure 5 (a)) but not in case of $d$ = 1.1, and 3.0 $\mu$m. As discussed in Refs. [43–46] the cracks are formed at the final drying stage of the droplet when the receding liquid tries to shrink the ring whereas the pinned particles obstruct such shrinkage. This results in stress inside the deposit and consequently induces cracks, explained in details in our previous work [38]. A bilayer of particles form in cases of $d$ = 1.1, and 3.0 $\mu$m since the ring height is almost double the particle diameter.

At $T_s$ = 60°C, the deposit patterns and the measured profiles are plotted in Figure 5 (b) and Figure 6 (b), respectively. The profiles show that the ring gets thinner in this case, and the deposit patterns show the presence of an inner deposit in this case. The ring width decreases by 86, 17, and 25 %, for $d$ = 0.1, 1.1, and 3.0 $\mu$m, respectively. At $d$ = 0.1 $\mu$m, a non-axisymmetric inner deposit forms while at $d$ = 1.1, and 3.0 $\mu$m, a ring-like axisymmetric inner deposit forms.

### 4.1.1  Mechanism of formation of the deposits

At $T_s$ =25°C, the evaporation-induced flow inside the droplet is radially outward, owing to the largest evaporation flux near the contact line, shown schematically in Figure 7 (a). The arrows normal to the liquid-gas interface represent the evaporation mass flux. Thus, the advection of the particles leads to the formation of the ring, as explained earlier by Deegan et al. [5]. In case of the substrate heating ($T_s = 60°C$), despite a heat loss by the latent heat of evaporation near the contact line, thermal energy is readily available from the substrate in this region, as discussed by Ristenpart et al. [17]. Since a thermal conduction resistance exists across the droplet thickness, the temperature on the liquid-gas interface near droplet apex is lesser than near the contact line. Due to these two factors, a thermal gradient appears across the liquid-gas interface. Consequently, it results in a surface tension gradient and a thermal Marangoni flow exists at the liquid-gas interface whose direction is from the contact line to the droplet apex [12–14]. Along with radially outward flow (bulk flow), it leads to an axisymmetric Marangoni recirculation inside the droplet that advects particles towards the inner region, as shown in Figure 7(b). The arrows on the liquid-gas interface in Figure 7(b) represent the variation of evaporation mass flux on a heated substrate, reported in Ref. [47]. The outward radial flow and the inward Marangoni flow, which are opposite in direction to each other, create a stagnation region near the contact line, shown schematically in Figure 7 (b). The existence of such a stagnation region was reported in Refs. [12–14]. Due to the curvature of the liquid-gas interface and the outward radial flow, some particles get deposited in the stagnation region, forming the outer ring [12, 14], while most of the particles are advected inward, due to the Marangoni recirculation. Therefore, the deposit, in the heated substrate case, is a thinner ring along with an inner deposit. The thinning of the ring occurs due to advection of particles towards the center of the droplet by the inward Marangoni flow. At $d = 0.1$ $\mu$m, the non-axisymmetric inner deposit is formed due to early depinning of the contact line with a stick-slip motion from random direction and the particles are dumped asymmetrically in the wetted area during the final dry out stage of evaporation. While at $d = 1.1$, and $3.0$ $\mu$m, the contact line is pinned throughout the evaporation, and the inward advection of the particles due to the Marangoni recirculation forms a ring-like axisymmetric inner deposit.

### 4.2 Non-uniform temperature substrate

Figure 8 (a), (b), and (c) show dried patterns of the deposit obtained by optical microscopy for the particle size of $d = 0.1$, $1.1$, and $3.0$ $\mu$m, respectively. Two runs are shown for each case, and the columns represent different cases of $dT/dX = 1.7$, $2.8$, and $4.2°C/mm$ and the left and right side of each case was at lower ($T_L$) and higher temperature ($T_H$), respectively. The zoom-in views of the ring on the $T_L$ and $T_H$ sides are shown for each case in Figure 8. The corresponding ring profiles measured using optical profilometer for the first run for $d = 0.1$, $1.1$, and $3.0$ $\mu$m for the different $dT/dX$ are plotted in Figure 9 (a), (b), and (c), respectively.

For $d$ = 0.1 $\mu$m, the ring width is wider on the $T_L$ side as compared to the $T_H$ side for all the cases of d$T$/d$X$ as observed qualitatively in Figure 8 (a) and quantitatively in Figure 9 (a). With an increase in d$T$/d$X$, the ring width on the $T_L$ side increases from 50 $\mu$m to 76 $\mu$m whereas the ring width on the $T_H$ side decreases from 44 $\mu$m to 8 $\mu$m. The ring heights are also larger on the $T_L$ side as compared to the $T_H$ side and increase with the increase in the d$T$/d$X$. Thus, the mass of particles is smaller on $T_H$ side as compared to that on the $T_L$ side. Cracks on the ring were recorded in this case, as seen earlier in case of the uniform temperature substrate.

In contrast, for $d$ = 1.1, and 3.0 $\mu$m, the ring width is wider on the $T_H$ side as compared to the $T_L$ side for all cases of d$T$/d$X$ as observed qualitatively in Figure 8 (b), and (c), and quantitatively in Figure 9 (b), and (c), respectively. In general, the ring width shows a slight increase and decrease for $T_H$ and $T_L$ side, respectively, with an increase in d$T$/d$X$. The ring height remains almost same for $d$ = 1.1 $\mu$m with an increase in d$T$/d$X$ while it increases for 3.0 $\mu$m. Overall, with the increase in d$T$/d$X$, the mass of particles in the ring is larger on the $T_H$ side as compared to that on the $T_L$ side. The mass of inner deposits on the $T_H$ side for $d$ = 1.1, and 3.0 $\mu$m increases with an increase in the d$T$/d$X$, as observed qualitatively in Figure 8 (b) and (c), respectively. The ring-like axisymmetric inner deposit was also recorded for $d$ = 1.1 $\mu$m with the intensity of the inner deposit larger towards the $T_H$ side, as observed qualitatively in Figure 8 (b).

### 4.2.1 Effect of particle size and temperature gradient on contact line motion

Time-varying droplet shapes recorded using side-visualization at d$T$/d$X$ = 4.2°C/mm for $d$ = 0.1, 1.1 and 3.0 $\mu$m are plotted in Figure 10 (a), (b), and (c), respectively. The respective movies S1, S2 and S3 are provided in the supporting information. In order to quantify the contact line motion in these measurements, the displacements of the contact line from low ($T_L$) and high ($T_H$) temperature sides are denoted as $X_L$ and $X_H$, respectively, and are schematically shown in Figure 11(a).

The contact line displacements for the three cases of particles size are compared in Figure 11(b) at d$T$/d$X$ = 4.2°C/mm. The contact line on $T_H$ side depins after 25 s for $d$ = 0.1 $\mu$m, as seen by the graph of $X_H$ (red stars). The contact line at the $T_L$ side is pinned for the maximum time of evaporation, as shown by the graph of $X_L$ (blue stars) in Figure 11 (b). These trends are qualitatively confirmed in Figure 10 (a). The jump in the contact line position from the $T_H$ side, corresponding to the depinning, is observed at around 40 s (confirmed in movie S1). The contact line exhibits stick-slip motion after this time. By contrast, in the cases of $d$ = 1.1 and 3.0 $\mu$m, the contact line remains pinned (see graphs of $X_L$ (blue squares and circles) and $X_H$ (red squares and circles)) for the maximum time during the evaporation and depins at around 70 s towards the end of evaporation (observed in Figure 10 (b), and (c) and Figure 11 (b)).

In Figure 11(c) for d$T$/d$X$ = 2.8ºC/mm and Figure 11(d) for d$T$/d$X$ = 1.7ºC/mm, similar trends of the contact line motion are noted for the three particle sizes considered, as discussed above for d$T$/d$X$ = 4.2ºC/mm. The contact line depins at around 80 s in case of d$T$/d$X$ = 2.8ºC/mm for $d$ = 0.1 $\mu$m particles, as plotted in Figure 11(c) (see also movie S4). Whereas, the contact line motion for d$T$/d$X$ = 1.7ºC/mm for $d$ = 0.1 $\mu$m is comparatively smoother in Figure 11(d) (see also movie S5). This is attributed to a slower evaporation in 1.7ºC/mm case as compared to 2.8ºC/mm and 4.2ºC/mm cases and the total evaporation time is 28% and 43% shorter, respectively. The contact line remains pinned for most of the evaporation for larger particles in cases of d$T$/d$X$ = 2.8ºC/mm as well as d$T$/d$X$ = 1.7ºC/mm, as shown in Figure 11(c) and Figure 11(d), respectively.

### 4.2.2 Effect of contact line depinning on the deposits

As discussed in section 4.2.1, the contact line depins in case of $d$ = 0.1 $\mu$m particles from the hot side of the substrate ($T_H$). The deposit patterns obtained for two cases of the temperature gradient, d$T$/d$X$ = 4.2ºC/mm and 2.8ºC/mm are shown in Figure 12(a) and (b), respectively. The insets in Figure 12 show a zoomed-in view of the deposit on $T_L$ side, top side and $T_H$ side. As discussed earlier, the deposits in both cases exhibit a smaller ring thickness at $T_H$ side while a larger ring thickness at $T_L$ side. Multiple thin lines of the deposits are observed in the microscopic images of the inner region of the deposit, as shown in the insets of Figure 12(a) and (b). These thin lines are the signatures of the stick-slip motion of the contact line, that depins from $T_H$ side for $d$ = 0.1 $\mu$m particles, as discussed earlier in section 4.2.1. Some large structures of the deposits in the insets of Figure 12 ($T_L$ side) are noted. These are the signatures of the dewetting and drying out of a residual thin liquid film containing particles on the substrate. These particles do not get advected towards the pinned or depinned contact line during the evaporation and remain trapped in the liquid film until at the end of the evaporation.

### 4.2.3 Mechanism of formation of the deposits

The preferential deposition of the particles on either $T_H$ or $T_L$ side with particle size and d$T$/d$X$ is explained as follows. The unidirectional temperature gradient on the substrate creates a temperature gradient on the liquid-gas interface. The temperature near the contact line region of $T_H$ side is larger as compared to the $T_L$ side, as expected. The measured temperature field using IR thermography on the evaporating droplet (provided in the supporting information,

Figure S1, movie S6 in supporting information), verifies the temperature gradient on the liquid-gas interface. The gradient causes thermocapillary Marangoni (interfacial) flow across the liquid-gas interface whose direction is from $T_H$ side to $T_L$ side. The non-uniform evaporation mass flux on the interface induces flow from $T_L$ side to $T_H$ side. This results in twin Marangoni recirculations in two equal halves of the droplet that are separated by a plane aligning along the direction of the imposed temperature gradient. The flow in this plane is schematically shown in Figure 13. Such twin recirculations in a pure water droplet on a non-uniformly heated substrate was measured by PIV technique in Ref. [30]. The two characteristics flow (bulk and interfacial) create stagnation regions on both $T_H$ and $T_L$ side contact line where the advected particles get trapped. The existence of a stagnation region on a uniformly heated substrate has been explained earlier in section 4.1.1. However, the size of the stagnation region on the two sides is not the same on a non-uniformly heated substrate (Figure 13). Owing to the larger evaporation flux on $T_H$ side, the flow towards the $T_H$ side would be stronger, creating a larger stagnation region on the $T_H$ side as compared to that on the $T_L$ side. At $d = 0.1 \mu$m, the depinning of the contact line occurs from the $T_H$ side leading to the stick-slip motion of the contact line towards to $T_L$ side, which advects particles from $T_H$ to $T_L$ side and thereby increases the ring dimensions on the $T_L$ side (Figure 13). However, at $d = 1.1$ and $3.0 \mu$m, the contact line is pinned for the maximum time of the evaporation, and a larger number of particles are deposited on the $T_H$ side in the larger stagnation region, as shown in Figure 13.

In order to explain the depinning and pinning of the contact line in case of smaller and larger particles, respectively, a mechanistic model (section 3) is employed. The dimensionless net force ($\Delta F^*$) acting in the inward horizontal direction on the particle nearest to the contact line (eq. 1) is plotted as a function of particles size in Figure 14 for different cases of d$T$/d$X$ imposed on the substrate. It can be noted that for smaller particles ($d = 0.1 \mu$m), $\Delta F^* > 0$, i.e., contact line depins for all cases of d$T$/d$X$, which signifies that the surface tension force overcomes the friction and drag forces together. By contrast, for larger particles ($d = 1.1$ and $3.0 \mu$m), $\Delta F^* < 0$, i.e. contact line pins for all cases of d$T$/d$X$, implying that the surface tension forces do overcome the friction and drag forces. In case of the smaller particles, the depinning occurs on the hot side since a larger intensity of the Marangoni flow advects particles from the hot side to the inner region and the number of particles stacked near the contact line is smaller on the hot side as compared to that on the cold side. Eq. 1 shows that for a smaller number of stacked particles ($n$), $\Delta F^* > 0$ or vice-versa. Overall, the model predictions are consistent with the measurements and this mechanistic model helps to explain the measurements.

## 4.3 Regime map

Finally, a regime map is plotted to classify the deposition patterns on a temperature gradient ($dT/dX$) - particles size ($d$) plane in Figure 15. At $dT/dX = 0$, i.e. if the substrate is non-heated or heated uniformly, the ring width is uniform. Thus, the first regime is called as *uniform ring width*. With an increase in the $dT/dX$, the ring width gets thicker on the $T_L$ and $T_H$ side for smaller and larger particles, respectively. The dashed lines demarcate the three regimes qualitatively, as shown in Figure 15. Our measurements show that for the evaporation on a non-uniformly heated substrate, there exists a critical $dT/dX$ at which ring width becomes non-uniform. Beyond this $dT/dX$, there also exists a critical particle size below which the contact line depins from the $T_H$ side and stick-slip motion occur, which advects the particles towards the $T_L$ side. Since the ring width is larger on the $T_L$ side, this regime is *thicker ring on lower temperature side*. For the larger particles, the contact line is pinned and a larger number of particles deposit on the $T_H$ side due to the asymmetric Marangoni recirculation. This results in a larger ring width on $T_H$ side and this regime is *thicker ring on higher temperature side*.

## 5 Conclusions

The deposit profile and pattern obtained after the evaporation of a sessile water droplet containing polystyrene colloidal particles on a non-uniformly heated glass substrate have been investigated. Specifically, the effects of the temperature gradient and particles size on the deposit pattern and ring dimensions were investigated. The time-varying droplet shapes from the side and the temperature of the liquid-gas interface from the top were recorded using high-speed visualization and infrared thermography, respectively. The dried patterns were visualized under an optical microscope and quantified using an optical profilometer. The following temperature gradients were imposed across the substrate, $dT/dX = 1.7$, 2.8, and 4.2ºC/mm, and the following particles diameter were considered, $d = 0.1$, 1.1, and 3.0 $\mu$m.

On the *uniform* heated substrate, the thinning of the ring with substrate temperature was recorded, which is attributed to an increase in axisymmetric Marangoni recirculation with the substrate temperature. On the *non-uniformly* heated substrate, the ring dimensions are smaller and larger on higher temperature side ($T_H$) for a smaller particle ($d = 0.1$ $\mu$m), and the larger particles ($d = 1.1$ and 3.0 $\mu$m), respectively. The displacement of the contact line on the $T_H$ and $T_L$ side reveals that at $d = 0.1$ $\mu$m, the contact line depins early from the $T_H$ side leading to the stick-slip motion of the contact line and twin asymmetric Marangoni recirculations advect more particles towards the $T_L$ side. However, at $d = 1.1$, and 3.0 $\mu$m, the contact line remains pinned on both the sides for a maximum duration of evaporation. The pinned contact line and the twin asymmetric Marangoni recirculations aid in trapping the particles in the larger stagnation region on the $T_H$ side contact line. Consequently, the ring width is larger on the $T_H$ side in this case. The variation of the temperature gradient across the substrate shows that the ring width significantly increases or decreases on $T_H$ or $T_L$ side, depending on the particle size. This can be attributed to a stronger Marangoni recirculation as the temperature gradient increases.

A mechanistic model is utilized to understand the role of particle size in pinning or depinning of the contact line. It was found that the depinning in case of the smaller particles occurs because the surface tension force on the particles stacked near the contact line overcomes the drag and friction force on them. Finally, a regime map to classify the deposition patterns on a temperature gradient - particles size (d) plane is proposed. Three regimes of deposit pattern are proposed. The first regime is *uniform ring width* i.e. well-documented coffee-ring effect, corresponds to uniformly non-heated or uniformly heated substrate. The second regime is *thicker ring on lower temperature side*, explained by asymmetric, twin Marangoni recirculations inside the evaporating droplet and contact line depinning in the presence of the smaller particles. The last and third regime is *thicker ring on higher temperature side*, attributed to the twin Marangoni recirculation and the contact line pinning with the bigger particles. Overall, the present results provide fundamental insights into ring formation on non-uniformly heated substrates and can help to design associated technical applications.

## 6   Supporting information

- Expressions for the forces acting on the particles at the contact line, values of the parameters used in the calculation of the forces and IR thermography of the evaporating droplet for $d$ = 0.1, 1.1, and 3.0 $\mu$m (PDF).

- Movie S1: Video of side-visualization of the evaporating droplet for the case of particle diameter, $d$ = 0.1 $\mu$m and $dT/dX$ = 4.2°C/mm.

- Movie S2: Video of side-visualization of the evaporating droplet for the case of particle diameter, $d$ = 1.1 $\mu$m and $dT/dX$ = 4.2°C/mm.

- Movie S3: Video of side-visualization of the evaporating droplet for the case of particle diameter, $d$ = 3.0 $\mu$m and $dT/dX$ = 4.2°C/mm.

- Movie S4: Video of side-visualization of the evaporating droplet for the case of particle diameter, $d$ = 0.1 $\mu$m and $dT/dX$ = 2.8°C/mm.

- Movie S5: Video of side-visualization of the evaporating droplet for the case of particle diameter, $d$ = 0.1 $\mu$m and $dT/dX$ = 1.7°C/mm.

- Movie S6: Video of IR thermography of the evaporating droplet for $d = 0.1$ $\mu$m at $dT/dX = 4.2$ºC/mm.

# 7  Acknowledgments

R.B. gratefully acknowledges financial support by a grant (EMR/2016/006326) from Science and Engineering Research Board (SERB), Department of Science and Technology (DST), New Delhi, India. AFM measurements of the glass slides were carried out in a central facility provided by IRCC, IIT Bombay.

# 9 Tables

Table 1. Measured initial static ($\theta_i$) and receding ($\theta_{rec}$) contact angles (in degrees) for different cases of particles diameter ($d$) and temperature gradient ($dT/dX$). The particles concentration is fixed at $c = 0.1$ %v/v. The uncertainty in these measurements is around ±2-3°.

| $d\overrightarrow{T/dX}$ | 1.7°C/mm | | 2.8°C/mm | | 4.2°C/mm | |
|---|---|---|---|---|---|---|
| $d$ ↓ | $\theta_i$ | $\theta_{rec}$ | $\theta_i$ | $\theta_{rec}$ | $\theta_i$ | $\theta_{rec}$ |
| 0.1 $\mu$m | 24° | 10° | 22° | 10° | 19° | 11° |
| 1.1 $\mu$m | 21° | 9° | 23° | 8° | 26° | 8° |
| 3.0 $\mu$m | 38° | 7° | 41° | 7° | 42° | 5° |

# 10 Figures

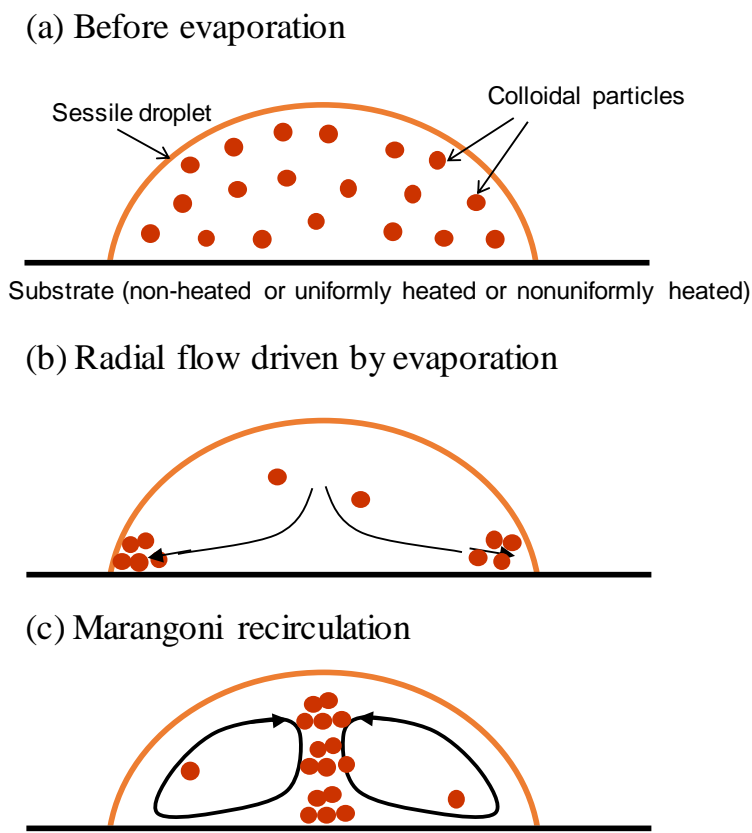

Figure 1: (a) Schematic of a sessile droplet loaded with colloidal particles on a solid surface. Coffee-ring effect on a non-uniformly heated substrate is studied in the present work. (b) Radially outward bulk flow advects all the particles to form a ring-like deposit. (c) Axisymmetric Marangoni flow recirculation in the presence of thermal gradient across the liquid-gas interface. The recirculation brings the particles from the contact line region to near the axis of symmetry. Adapted from Bhardwaj et al. [36]. Copyright, 2010, American Chemical Society.

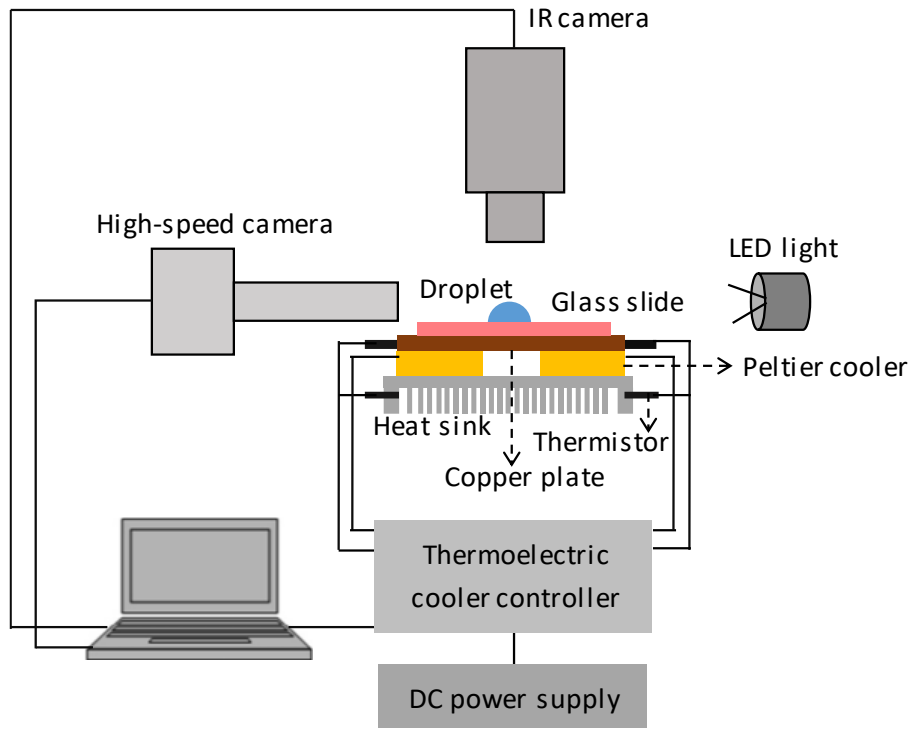

Figure 2. Schematic of the experimental setup used in the present study. Two Peltier coolers were used to impose a thermal gradient on the substrate. Infrared thermography and high-speed visualization were employed to map the temperature field on the liquid-gas interface and to record time-varying droplet shapes during the evaporation.

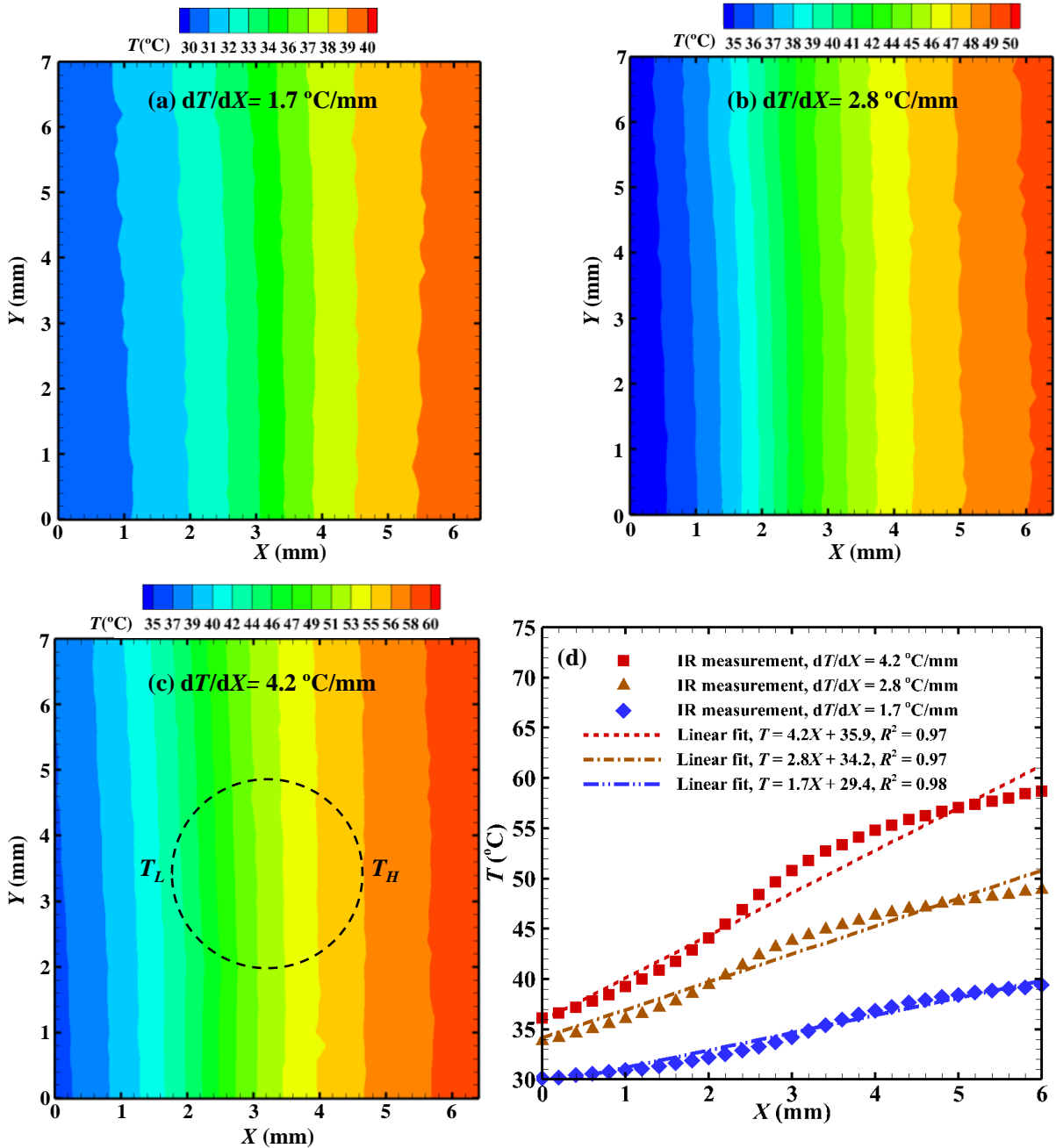

Figure 3. Isotherms plotted for different cases of imposed temperature gradient on the substrate (a) d$T$/d$X$ = 1.7°C/mm (b) d$T$/d$X$ = 2.8°C/mm (c) d$T$/d$X$ = 4.2°C/mm. The dotted circle shows the approximate location of the deposition of the droplet. (d) The temperature along $X$ at $Y$ = 3 mm is plotted for the three cases. Linear fits obtained for the measured data are shown with $R^2$ values.

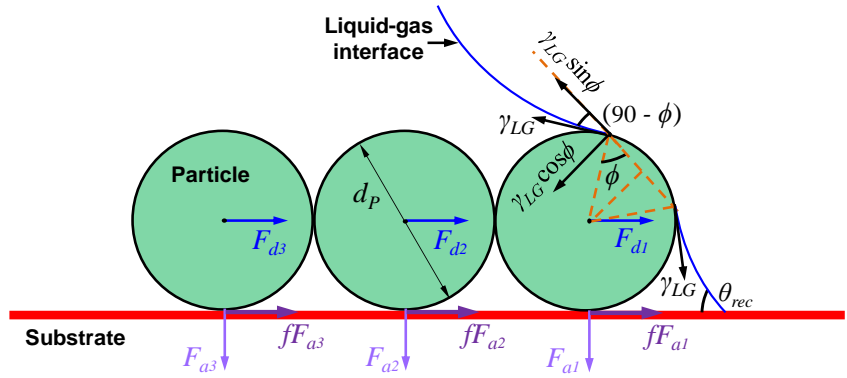

Figure 4. Schematic of particles stacked near the contact line and different forces acting on them. Surface tension force pulls the outmost particle inward while drag force pushes all particles outward towards the contact line. Adhesion force (van der Waals force) resist the motion of the particles. Adapted from Ref. [41], Copyright, 2017, Springer Nature and reproduced from Ref. [26], Copyright, 2018, American Chemical Society.

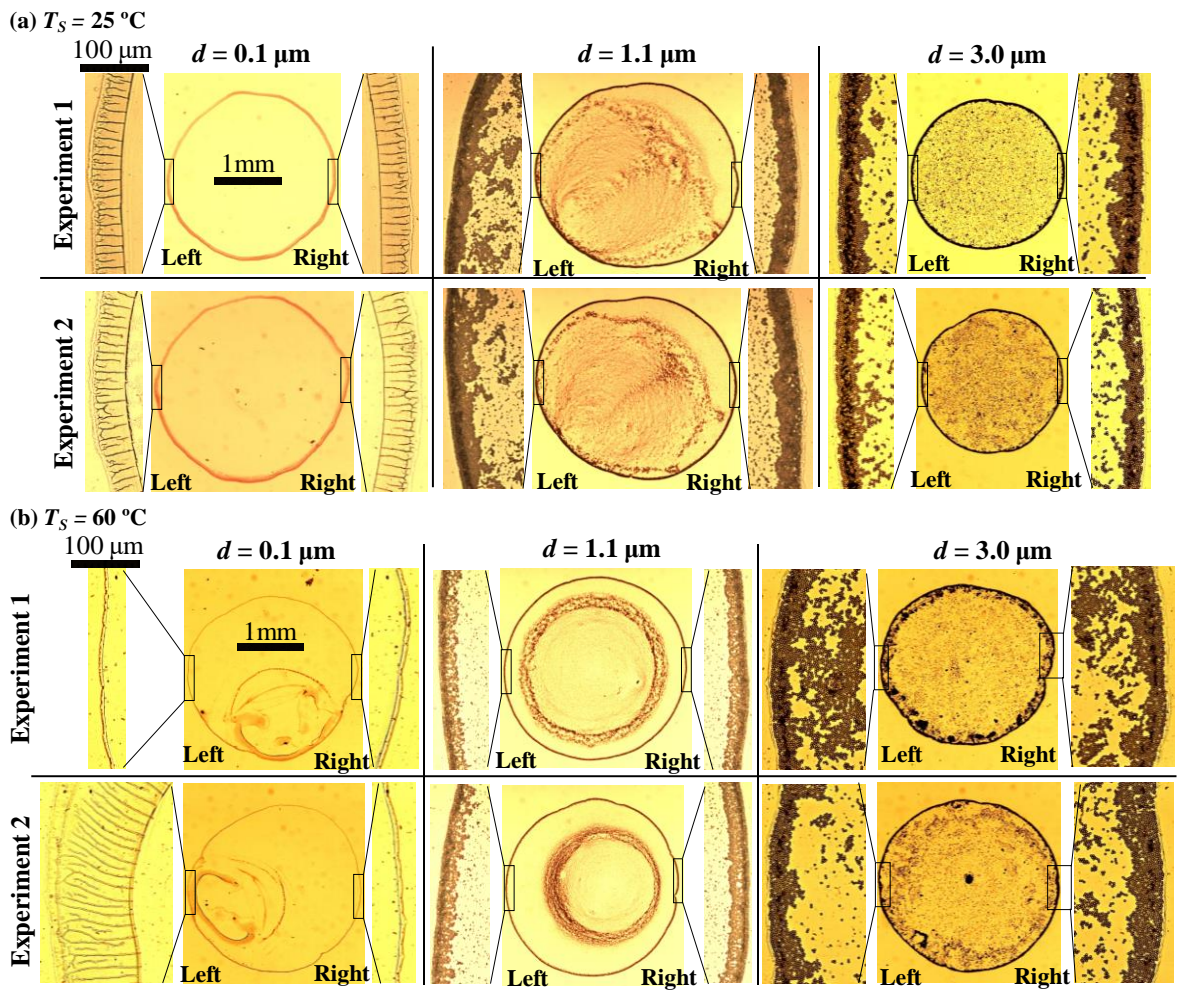

Figure 5. Deposit patterns obtained after evaporation of microliter aqueous droplet containing polystyrene particles of different particles sizes ($d$), keeping particles concentration same, $c = 0.1\%$ on a uniform non-heated or heated substrate (a) $T_s = 25°C$ (b) $T_s = 60°C$. Zoomed-in view on the left and right side of the ring is shown for each case.

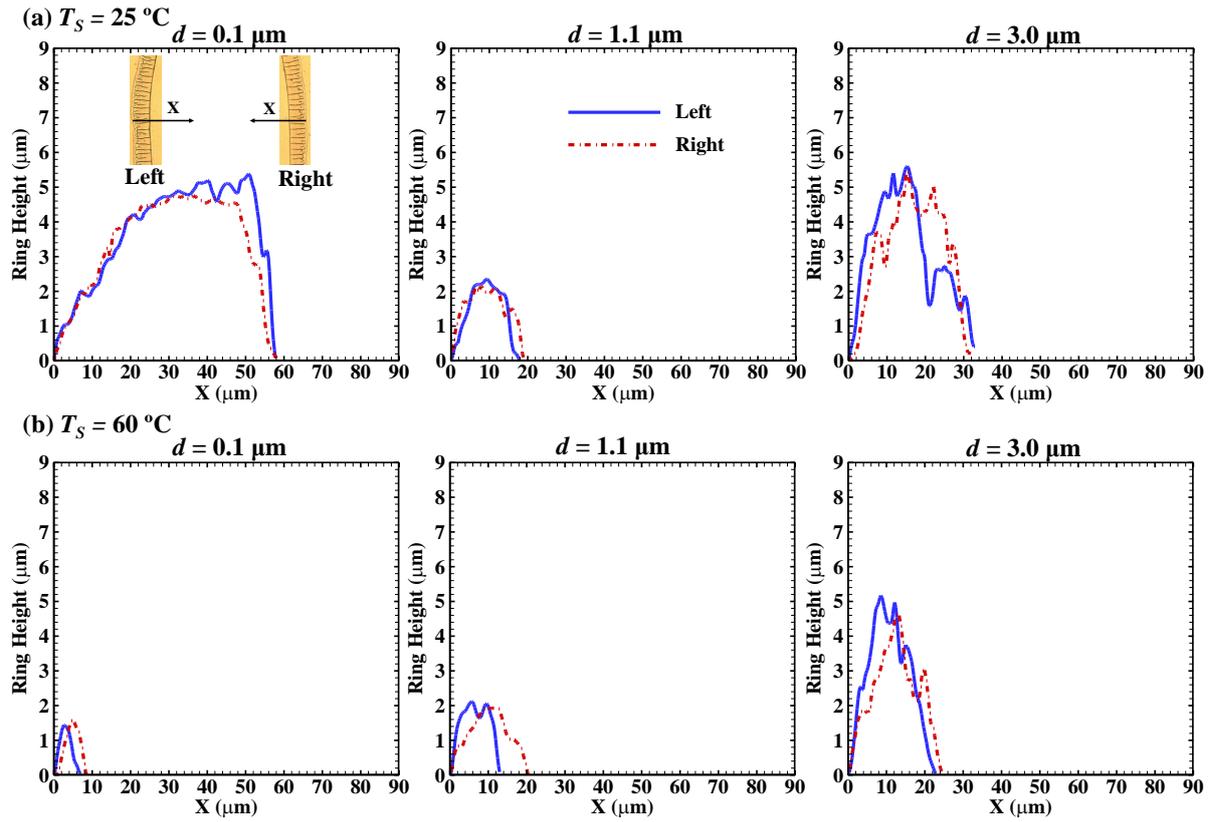

Figure 6. Measured ring profiles on the left and right side of the ring for different cases of substrate temperature (a) $T_s = 25°C$ (b) $T_s = 60°C$. Column represented data for three cases of particles size, $d = 0.1, 1.1, 3.0\,\mu m$, and particles concentration is 0.1% in all cases. X represents the radial position in the deposit (shown as the inset), and X = 0 is the ring periphery on both the left and right side.

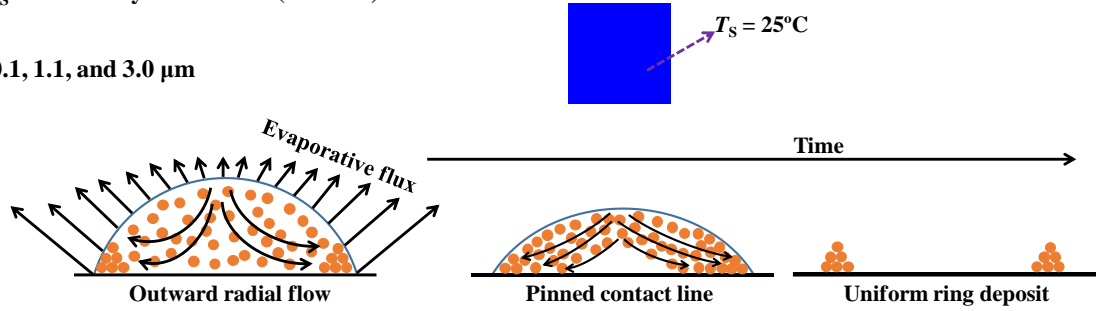
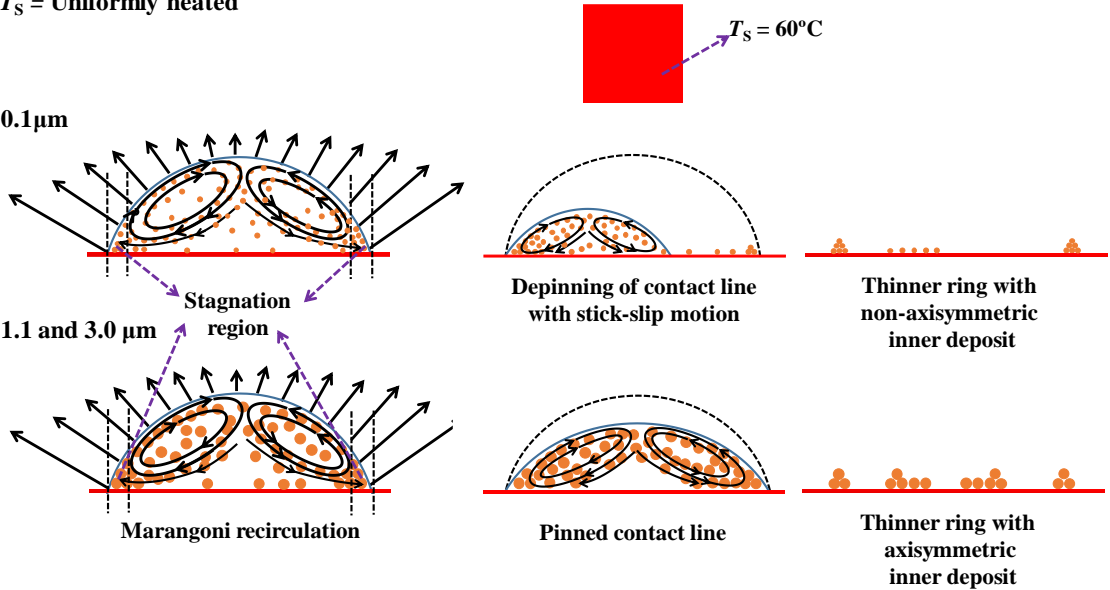

Figure 7. Mechanism of the deposit formation on a uniform temperature substrate. (a) Non-heated ($T_s = 25°C$) The arrows on the liquid-gas interface represent non-uniform, axisymmetric evaporation mass flux on the liquid-gas interface. A radial outward bulk flow advects particles to form a ring-like deposit. (b) Uniformly heated substrate ($T_s = 60°C$). An axisymmetric Marangoni recirculation develops due to themocapillary flow and radial flow in the bulk. In the case of smaller particles, the contact line depins and shows the signature of stick-slip motion while for the bigger particles the contact line remains pinned. The depinning occurs if the surface tension force on the particle at the contact line overwhelms the combined friction force between the particles and substrate and hydrodynamic drag force.

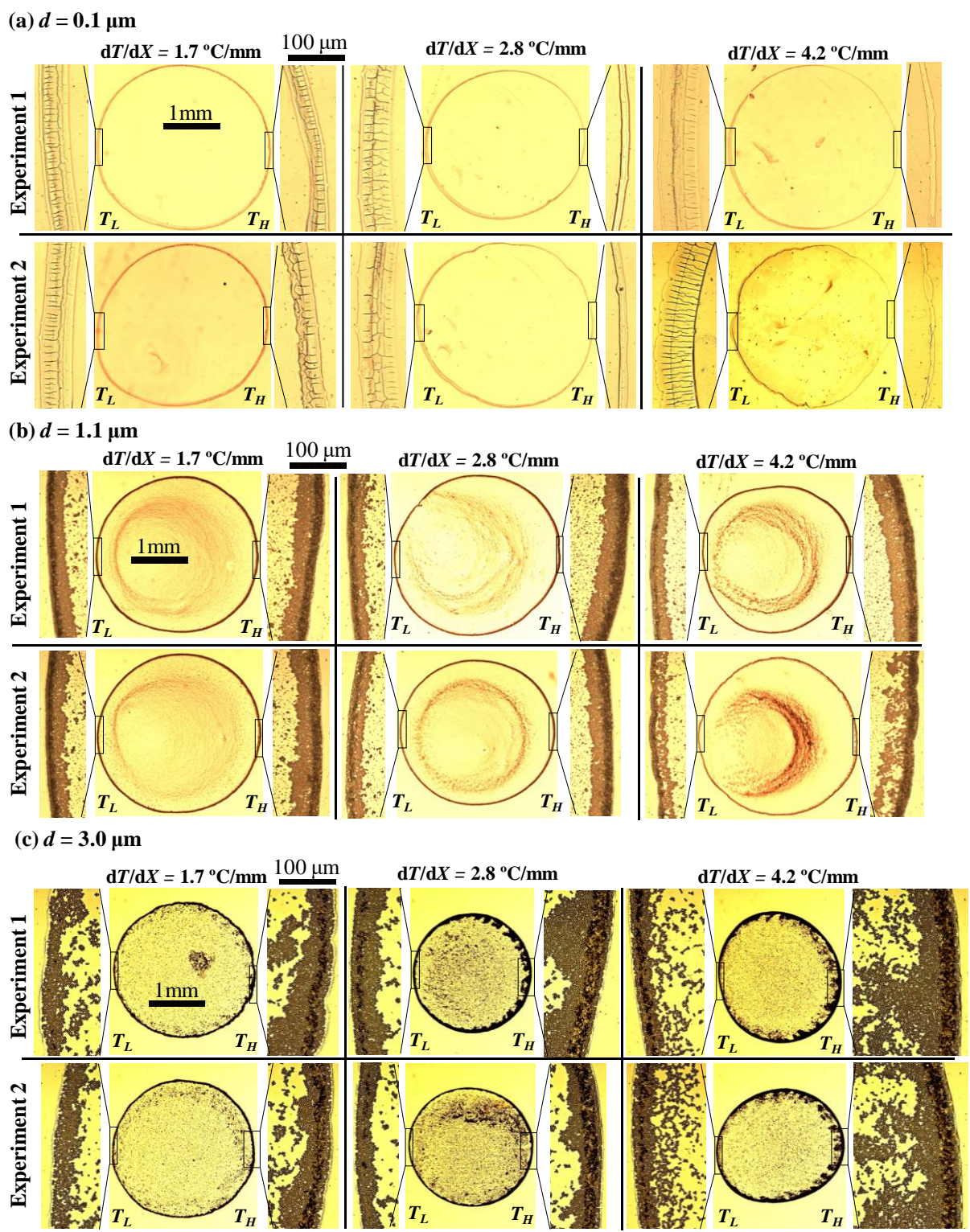

Figure 8. Deposit patterns obtained after evaporation of microliter aqueous droplet containing polystyrene particles of diameter, (a) $d = 0.1$ μm, (b) $d = 1.1$ μm, and (c) $d = 3.0$ μm, keeping particles concentration same, $c = 0.1$ % v/v. Column represent different cases of imposed temperature gradient, $dT/dX = 1.7, 2.8, 4.2$°C/mm. Two runs for each case are shown in different rows. $T_L$ and $T_H$ represent the lower and higher temperature side of the substrate, respectively.

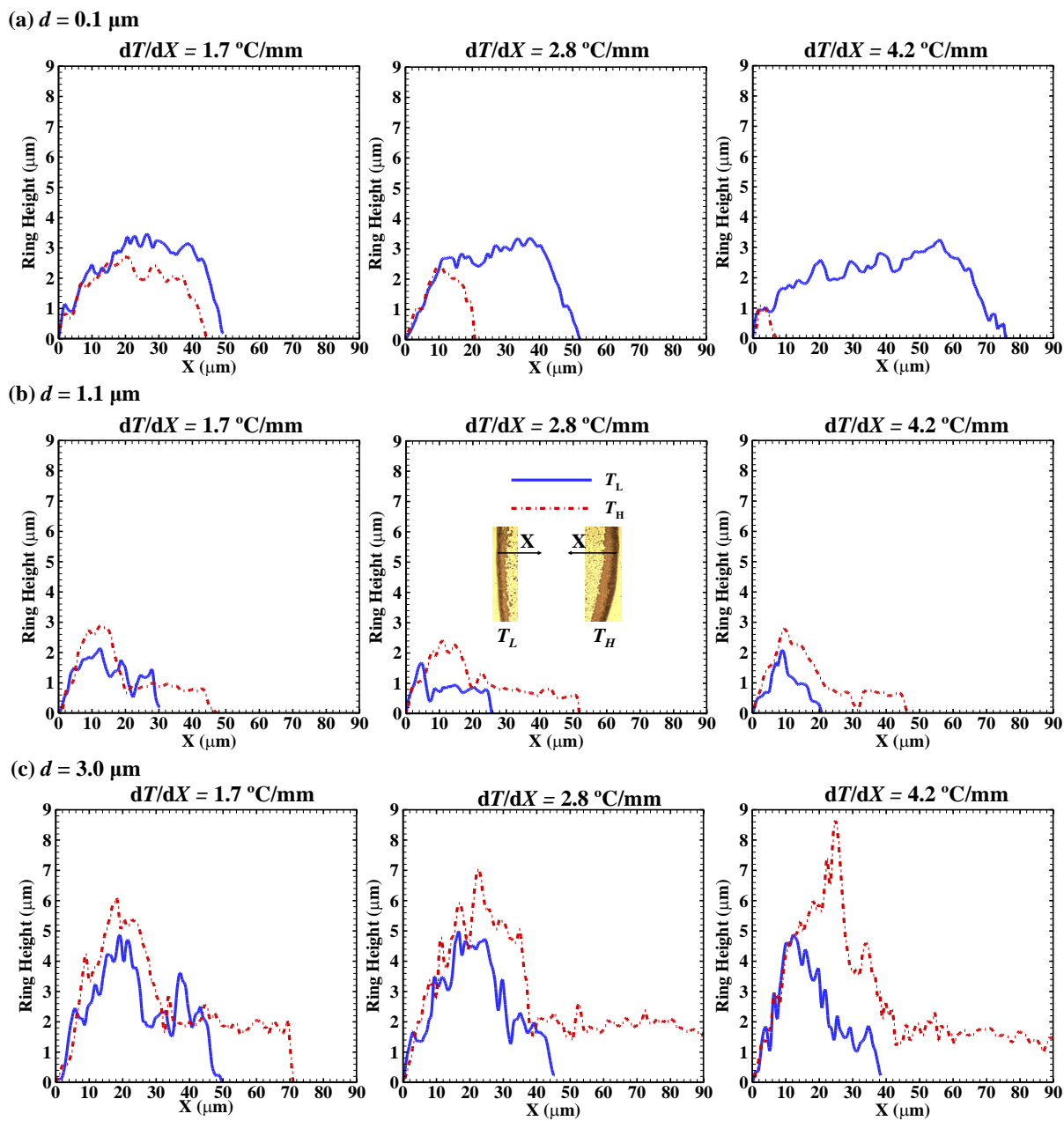

Figure 9. Measured ring profiles on the $T_L$ and $T_H$ side of the ring for different cases of particles size (a) $d = 0.1$ μm (b) $d = 1.1$ μm (c) $d = 3.0$ μm, keeping particles concentration same, $c = 0.1$ % v/v. Columns represented data for three cases of temperature gradients, $dT/dX = 1.7, 2.8, 4.2$ °C/mm. $T_L$ and $T_H$ represent the lower and higher temperature side of the substrate, respectively. X represents the radial position in the deposit (shown as the inset), and X = 0 is the ring periphery on both the $T_L$ and $T_H$ side.

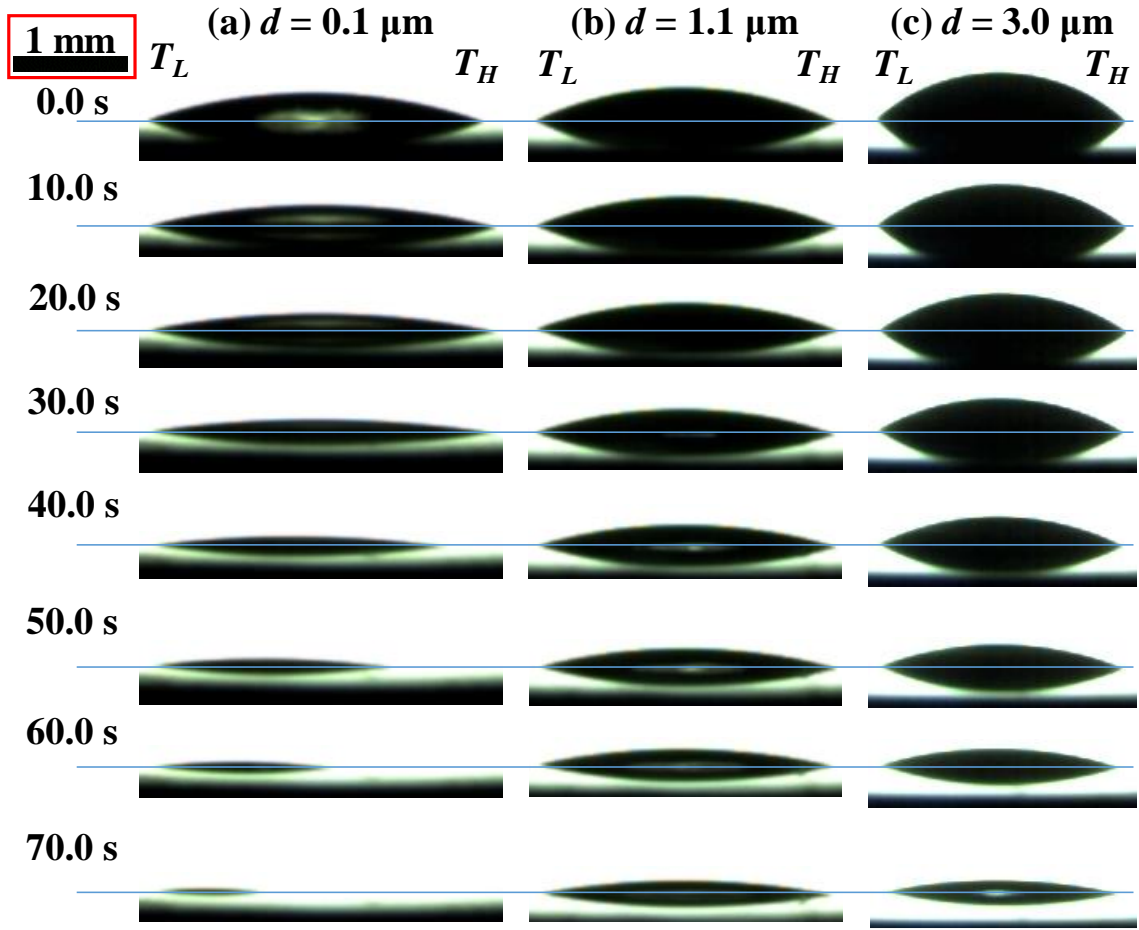

Figure 10. Time-sequence of the images obtained by side-visualization during the evaporation of a microliter colloidal suspension on a glass substrate, imposed with a temperature gradient of, $dT/dX = 4.2°C/mm$. A horizontal blue line in all frames represents the solid-gas or liquid-solid interface. $T_L$ and $T_H$ represent the lower and higher temperature side of the substrate, respectively. Three cases of particle sizes are plotted (a) $d = 0.1$ $\mu m$, (b) $d = 1.1$ $\mu m$, and (c) $d = 3.0$ $\mu m$. Movies S1, S2, and S3 for these cases, respectively, are provided in the supporting information.

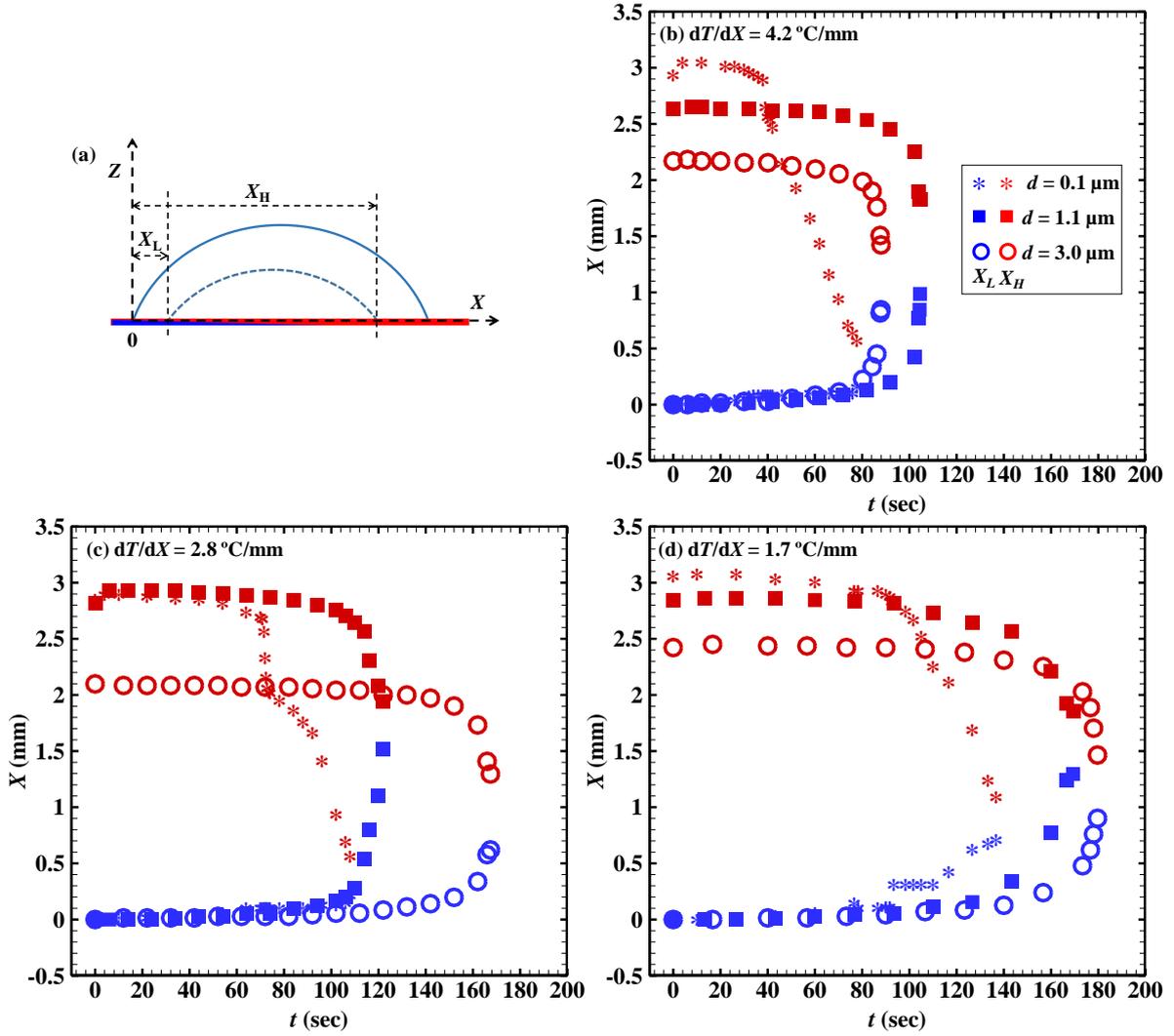

Figure 11. Time-variation of the displacements of the contact line from lower ($T_L$) and higher temperature ($T_H$) sides of the substrate. $X_L$ (blue symbols) and $X_H$ (red symbols) denote the displacements from $T_L$ and $T_H$ sides, respectively, as schematically shown in (a). Three cases of d$T$/d$X$ are plotted (b) 1.7°C/mm (c) 2.8°C/mm and (d) 4.2°C/mm. In each frame, the displacements, $X_L$ and $X_H$, are compared for three different particles sizes. For $d$ = 0.1 $\mu$m, depinning of contact line occurs from $T_H$ (red color) side in all cases of d$T$/d$X$, leading to the stick-slip motion of the contact line. For $d$ = 1.1, and 3.0 $\mu$m, the contact line is pinned for the maximum duration of the evaporation.

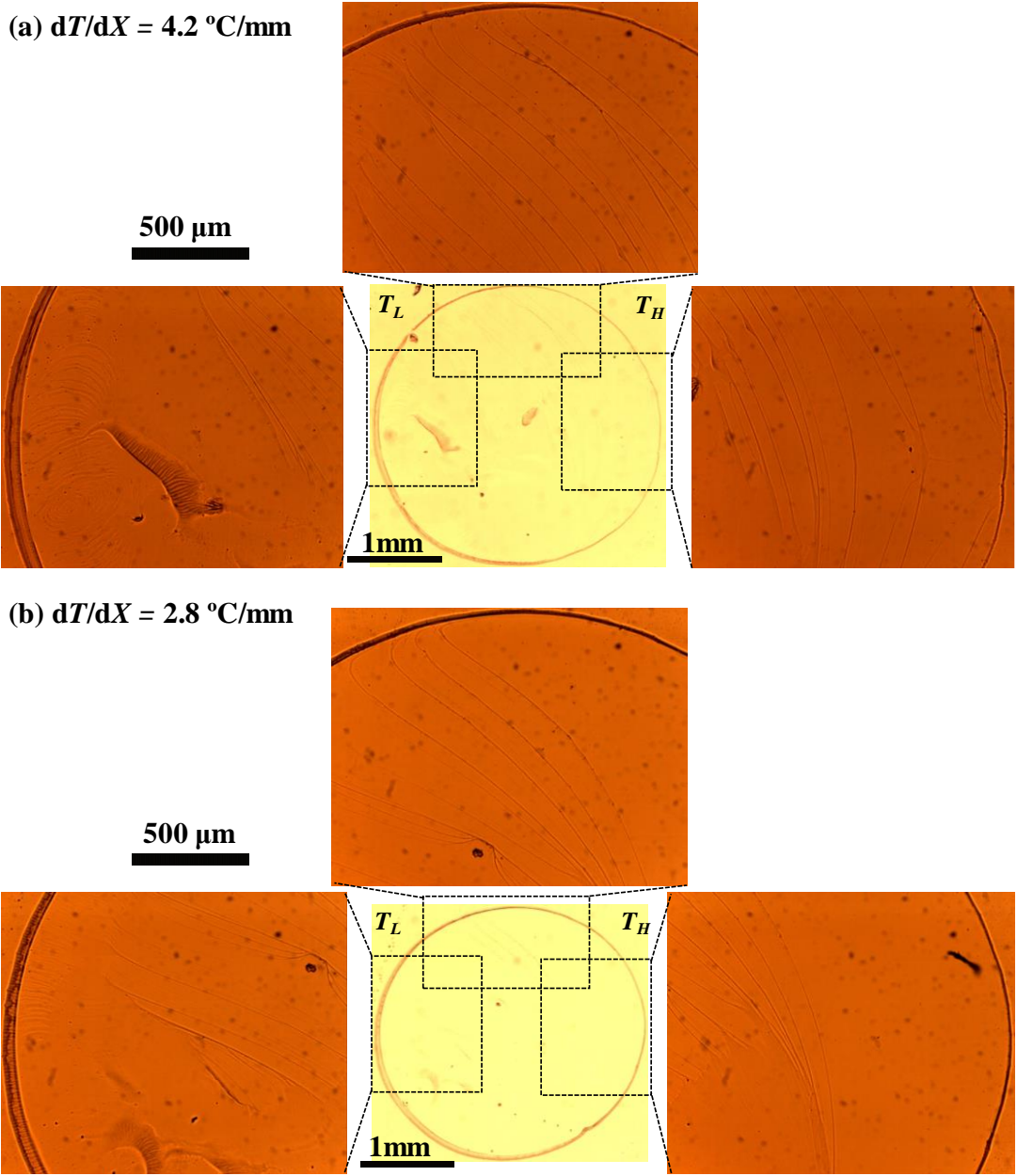

Figure 12. Deposit patterns obtained after evaporation of a microliter aqueous droplet containing polystyrene particles of diameter, $d = 0.1$ μm. $T_L$ and $T_H$ represent the lower and higher temperature side of the substrate, respectively. Insets show a zoomed-in view of inner deposits for three sides, namely, $T_L$ side, top side and $T_H$ side. Two cases of temperature gradient are shown: (a) $dT/dX = 4.2$°C/mm and (b) $dT/dX = 2.8$°C/mm. The patterns in the inset show the signature of the stick-slip motion of the contact line that depins from the hot side ($T_H$).

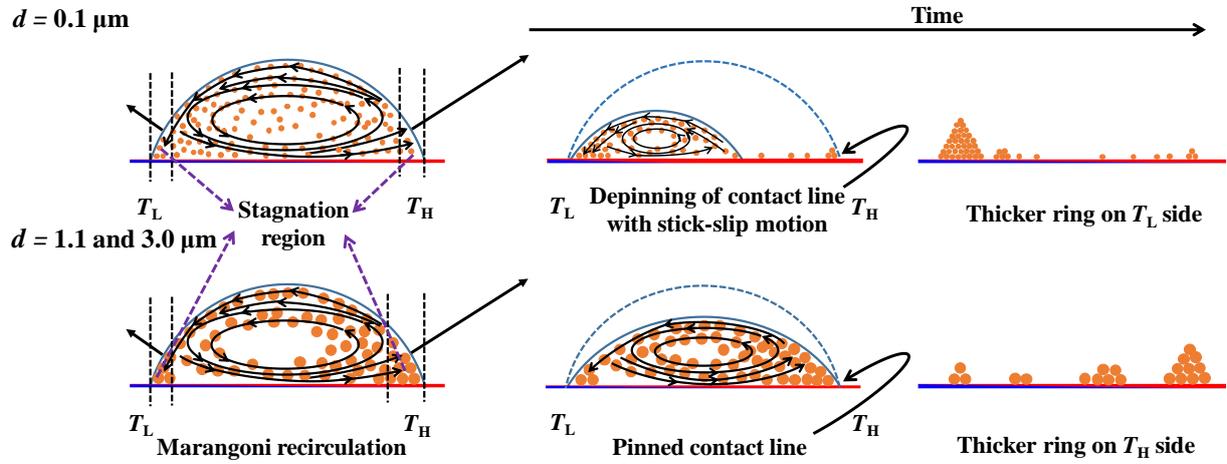

Figure 13. Mechanism of the deposit formation on a non-uniform temperature substrate for smaller ($d = 0.1$ $\mu$m) and larger particles ($d = 1.1$ and $3.0$ $\mu$m). $T_L$ and $T_H$ represent the lower and higher temperature side of the substrate, respectively. The arrows on the liquid-gas interface represent non-uniform and asymmetric evaporation mass flux. The flux is larger on the hotter side due to non-uniform heating of the substrate. The unidirectional temperature gradient on the substrate creates a temperature gradient on the liquid-gas interface leading to the thermocapillary Marangoni (interfacial) flow from $T_H$ side to $T_L$ side. The non-uniform flux together with the Marangoni flow on the liquid-gas interface induces bulk liquid flow from $T_L$ side to $T_H$ side that eventually develops as twin asymmetric Marangoni recirculations inside the droplet. In the case of smaller particles, the contact line depins and shows the signature of stick-slip motion while for the bigger particles the contact line remains pinned. Consequently, the final deposit has a larger ring thickness on the cold side for the smaller particles while a larger ring thickness results on the hot side for the bigger particles.

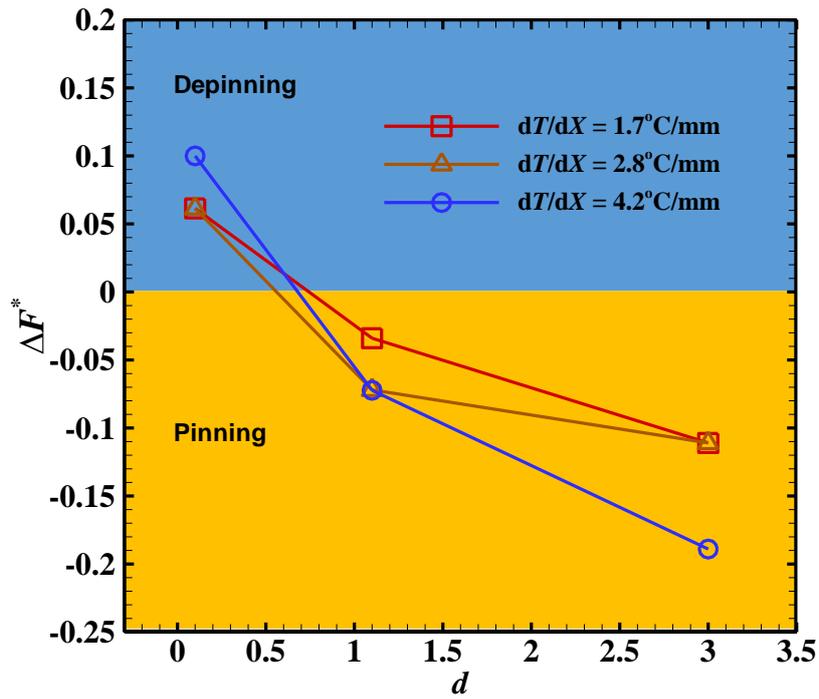

Figure 14. Dimensionless net force ($\Delta F^*$) in the horizontal direction acting on the particles stacked at the contact line is plotted as a function of particles size ($d$). Different cases of temperature gradients imposed on the substrate ($dT/dX$) are compared. Sign on the net force determines if the contact line pins ($\Delta F^* < 0$) or depins ($\Delta F^* > 0$).

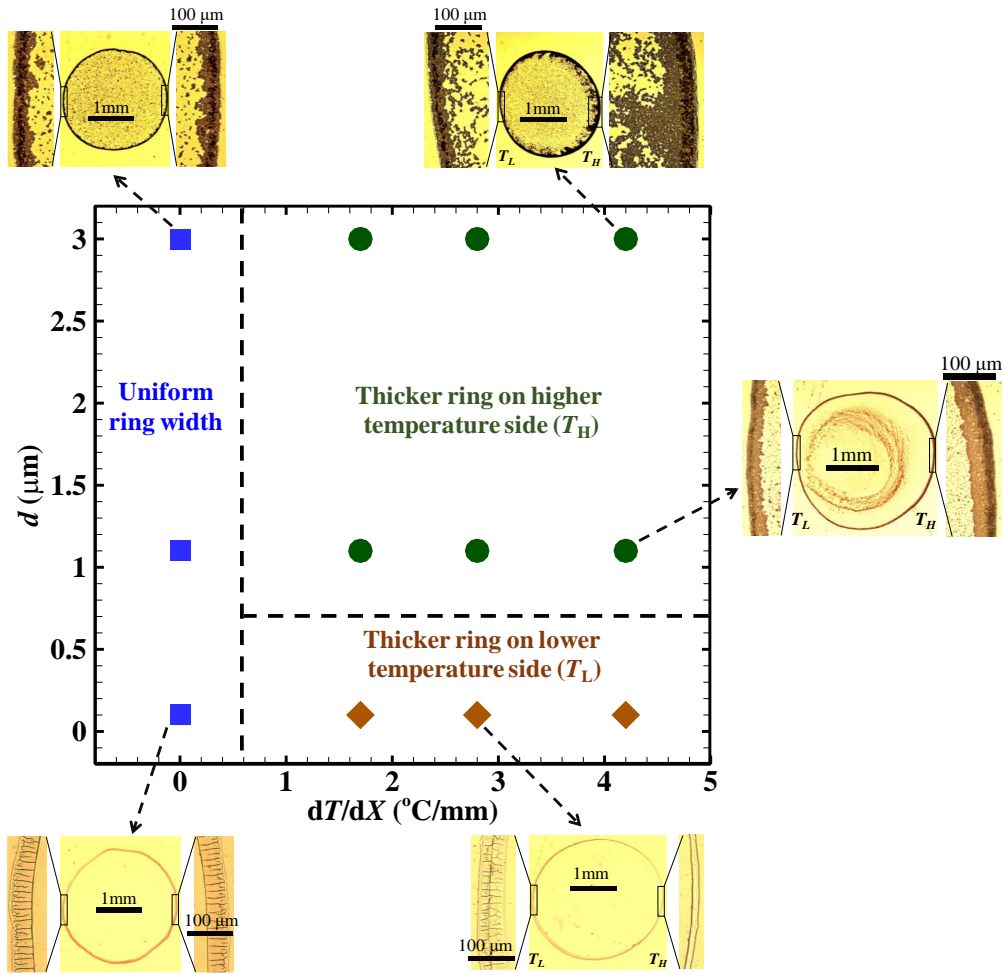

Figure 15. Regime map classifying different deposition patterns obtained as a function of temperature gradient imposed on the substrate (d$T$/d$X$) and particles size ($d$). Dashed lines are a guide to the eye and are plotted to demarcate the regimes. Insets corresponding to the different regimes are also shown with scale bars. Three regimes are found, namely, uniform ring width, thicker ring on lower temperature side and thicker ring on higher temperature side. The first regime i.e. coffee-ring effect is well-established in the literature while the regimes with non-uniform ring width are the contribution of the present work. The second regime is explained by asymmetric Marangoni recirculation inside the evaporating droplet and contact line depinning in the presence of the smaller particles. The third regime is the result of the recirculation and the contact line pinning with the bigger particles.